\documentclass[letterpaper,conference]{IEEEtran}
\usepackage{color}
\usepackage{epsfig}
\usepackage{amsmath}
\usepackage{amsfonts}
\usepackage{amsthm}
\usepackage{amssymb}
\usepackage{graphics}
\usepackage{graphicx}
\usepackage{float}

\newtheorem{algorithm}{Algorithm}
\let\olddefinition\algorithm
\renewcommand{\algorithm}{\olddefinition\normalfont}

\newcommand{\Z}{\mathbb{Z}}
\newcommand{\R}{\mathbb{R}}

\title{\LARGE \bf Multilevel MIMO Detection with Deep Learning}

\author{Vincent Corlay$^{\dagger,*}$, Joseph J. Boutros$^{\ddagger}$, Philippe Ciblat$^{\dagger}$, and Lo\"ic Brunel$^*$\\
$^{\dagger}$ Telecom ParisTech, 46 Rue Barrault, 75013 Paris, France, Tel. +33 1 4581 7728, 
v.corlay@fr.merce.mee.com\\
$^{\ddagger}$ Texas A\&M University, Doha, Qatar, $^*$Mitsubishi Electric R\&D, Rennes, France
}

\begin{document}

\maketitle

\begin{abstract} A quasi-static flat multiple-antenna channel is considered.
We show how real multilevel modulation symbols can be detected via deep neural networks.
A multi-plateau sigmoid function is introduced. Then, after showing
the DNN architecture for detection, we propose a twin-network
neural structure. Batch size and training statistics for efficient learning are investigated. Near-Maximum-Likelihood performance
with a relatively reasonable number of parameters is achieved. 
\end{abstract}

\section{Introduction}

In 2012  Alex Krizhevsky and  his team presented a  revolutionary deep
neural network  (DNN) in the  ImageNet Large Scale  Visual Recognition
Challenge~\cite{NIPS2012}.  The network  largely outperformed  all the
competitors. This event  triggered not only a revolution  in the field
of computer  vision but has  also affected many  different engineering
fields, including the field of digital communications. 
In our specific area of  interest, a  lot of new  studies were published  on machine learning for coding and communication theory since 2016. \\

In our work, we  address  the  case  of  multilevel symbol  detection  on  multiple-input multiple-output (MIMO)
channels via deep neural networks. 
There exist many algorithms to perform MIMO detection, whose performance ranges from optimal to highly suboptimal.
A first category of decoders includes sphere decoding methods based on lattice points enumeration and radius adaptation.
The complexity of sphere decoding is clearly less prohibitive than an exhaustive search and is polynomial in the dimension
for small dimensions. Detection based on sphere decoding is quasi-optimal and is very competitive in terms of number of operations for dimensions
less than 32 (up to 64 for non-dense MIMO lattices), however it cannot be parallelized because of its sequential nature.
Furthermore, the dynamic tree structure of sphere decoding makes it hardware-unfriendly.

In a second category we find linear receivers: the zero-forcing (ZF) detector and the minimum mean squared error (MMSE) detector.
Finally, a non-exhaustive list of decoders having performance somewhere between these two categories includes: the decision feedback-equalizer (DFE),
the K-best sphere decoder, message passing methods (e.g. belief propagation, approximate message passing, expected propagation) and semidefinite relaxation.
While some of these algorithms are near-optimal in specific settings, their performance are largely degraded when these specific conditions are not respected. 
As a result, the problem of finding hardware-friendly low-complexity methods exhibiting near-optimal performance in most settings remains open.
Neural network based implementation could offer new solutions.\\

MIMO detection with neural networks has already been investigated by several research groups.
In~\cite{Samuel2017} \cite{Samuel2018}, the  quadratic  form  of  the MIMO channel  is  used to build the network. 
In \cite{Tan2018} \cite{Liu2018} \cite{Imanishi2018} \cite{He2018} sub-optimal message passing iterative MIMO decoders are improved with the approach introduced in \cite{Hershey2014} \cite{Nachmani2016}. 
The main idea of these studies is to unfold the underlying graph used by an iterative algorithm to get improvement via learning.
Simulations show that in most cases learning enhances the performance of the considered algorithm. 
Nonetheless, these results are almost never compared to optimal detection. It is therefore difficult to assess the real efficiency of such an approach.
Additionally, most studies consider binary inputs only.
In \cite{Samuel2018}, one-hot encoding is used to address the case of non-binary inputs.
Unfortunately, the number of output neurons increases significantly with the spectral efficiency making this solution impractical. 

\section{Problem settings and network used}

In this  paper we  use row  convention for  vectors and  matrices.  We
consider a  symmetric flat  quasi-static MIMO channel with  $n$ transmit antennas and $n$
receive antennas. Let $G$ be the  $n \times n$ matrix representing the
channel coefficients.  For simplicity, it is assumed that $G$ has real
entries.
Any complex matrix of size $n/2$ can be trivially transformed into a real matrix of size $n$.
Let $z  \in \Z^n$ be  the channel
input,  i.e., $z$  is  the uncoded  information  sequence.  The  input
message yields  the output  $y \in  \R^n$ via  the standard  flat MIMO
channel equation,
\vspace{-1mm}
\begin{align*}
y = \underset{x}{\underbrace{z \cdot G}} + \eta,
\end{align*}
where $\eta$ is a Gaussian vector with i.i.d. $\mathcal{N}(0, \frac{N_0}{2})$ components. 
The optimal decoder, also called Bayes decoder in the machine learning community, implements the maximum a posteriori (MAP) criterion. A near-optimal neural network detector should implement a function $f$ that approximates the MAP criterion.
\begin{align*}
f(y) \approx \underset{z \in \mathcal{M}}{\text{arg max}} \ P(z|y),
\end{align*}
where $\mathcal{M}$ is the finite MIMO constellation. 
In our settings, the MAP criterion is equivalent to finding the closest possible $x$, closest in the Euclidean sense, as expressed by the following equation:
\begin{align*}
\hat{z} = \underset{z \in \mathcal{M}}{\text{arg min}} \ ||y-\underset{x}{\underbrace{z.G}}||.
\end{align*}

Neurons in regular DNN include a non-linear activation function, such as the sigmoid function
and the rectified linear unit~\cite{Goodfellow2016}. 
In the sequel, the standard sigmoid function $\sigma(t) = 1/(1+e^{-t})$ is employed.

\subsection{Architecture}

\begin{figure}
\centering
\includegraphics[width=0.71\columnwidth]{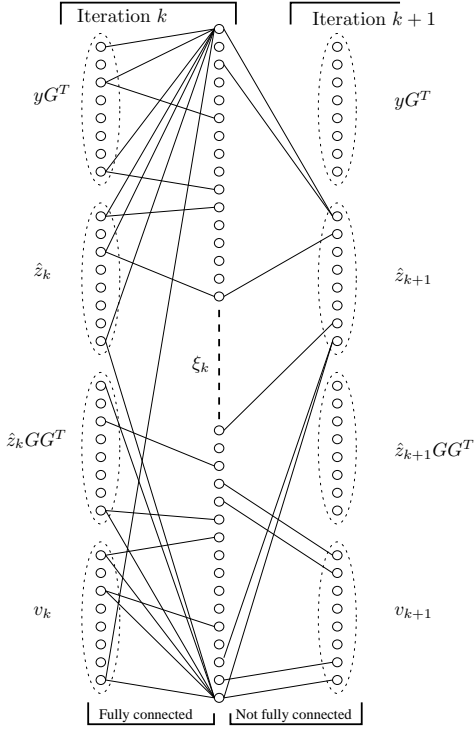}
\vspace{-2mm}
\caption{DNN architecture, two layers per iteration.}
\label{fig_DNNarchitecture}
\vspace{-5mm}
\end{figure}

In  \cite{Samuel2017}, the architecture of the  network is inspired from  the projected gradient descent:

\begin{align*}
\hat{z}_{i+1} &= \scriptstyle {\cal \prod } \left( \hat{z_{i}} - \eta \cdot \frac{\partial ||y-z.G||^2}{\partial z} \vert_{z=\hat{z}_{i}}  \right) = \scriptstyle {\cal \prod } \left( \hat{z_{i}} - 2 \eta \cdot yG^{T} + \eta \cdot \hat{z}_{i} G G^{T}   \right),
\end{align*}
where $\scriptstyle {\cal \prod}$ is a projection operator.
Our neural network embraces the same paradigm. It takes the form of an iterative algorithm where an estimate of the output is available after each iteration.
It is illustrated in Figure~\ref{fig_DNNarchitecture}.
A~generic iteration has two layers, as shown in the figure, where the network structure is derived
from the following matrix equations:
\begin{flalign*}
&\xi_{k} = \sigma_{c}\left( W^1_{1k}\hat{z}_{k} + W^2_{1k}yG^T +  W^{3}_{1k}\hat{z}_{k}GG^{T} + W^{4}_{1k}v_{k} + \text{bi}_{1k} \right),\\
&\hat{z}_{k+1} = \sigma_{c}\left( W_{2k} \xi_{k} + \text{bias}_{2k} \right), \ \ v_{k+1}=W_{3k}z_{k}+ \text{bias}_{3k}.
\end{flalign*}
In the expression of $\xi_{k}$, we can clearly recognize the terms used by the gradient descent, weighted by $W^i_1$ instead of $\eta$ (the two other terms are a hidden variable and a bias term commonly used in neural networks). 
The intuition behind this expression is that the network will learn specific learning rates $\eta$ for each iteration and each component. 
The operation performed between the $\xi$ layer and the next layer can be interpreted as the projection operator~$\scriptstyle {\cal \prod }$. The activation function used $\sigma_{c}$ is described in the next section.

In  \cite{Samuel2017}, the matter of how $\hat{z}_0$ should be initialized for the first iteration of the neural network is not discussed. We address and take advantage of this question in the section on the twin-network.

\subsection{The multilevel activation function }

The default approach to address a multi-class problem with neural networks is to use the so-called ``one-hot encoding". Namely, if the network should classify data between more than two categories, say $M$ categories, it will have $M$ output neurons where legal combinations of values are only the $M$ combinations with a single neuron equal to 1 and all the others equal to 0. Unfortunately, this approach implies a large amount of output neurons. In the network of Figure~\ref{fig_DNNarchitecture}, if each component of the input message $z$ can take $M$ levels, using one-hot encoding means having $n \times M$ output neurons (the neurons labeled $z_{k+1}$ in Figure~\ref{fig_DNNarchitecture}) instead of $n$ in the binary case. This implies a greater complexity as well as longer training.
\begin{figure}
\centering
\includegraphics[angle=270,width=0.45\columnwidth]{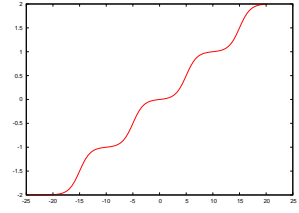}
\vspace{-1mm}
\caption{$\sigma_{c}(t)$ for 5-level integer symbols.}
\label{fig_custoSig}
\vspace{-5mm}
\end{figure}

To address this issue we introduce a novel activation function: we adapt the non-linearity in the output neurons to take into account non-binary symbols.
Our customized sigmoid function shall be defined as a sum of standard sigmoids,
\[
\sigma_{c}(t) = \sum_{i=1}^M \sigma(t-\tau_i) + A,
\]
where $\tau_i$ are sigmoid shifts and $A$ is an overall translation.
As an example, for $z \in \{-2,-1,0,1,2\}^n$ ($M=5$), the customized sigmoid is taken to be
$\sigma_{c}(t) = \sigma(t+15) + \sigma(t+5) + \sigma(t-5) + \sigma(t-15)+ \sigma(t-25) -2$, as
depicted on Figure~\ref{fig_custoSig}. 

\subsection{The twin-network}

\begin{figure}
\centering
\includegraphics[width=0.85\columnwidth]{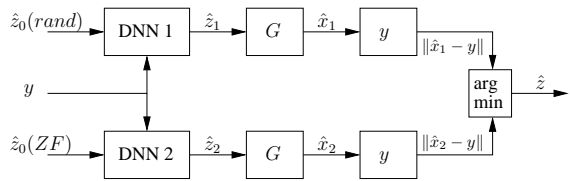}
\caption{Block representation of the neural system. \label{fig_Blockrepre}}
\vspace{-5mm}
\end{figure}

To further improve our system, we considered the paradigm of a random forest~\cite{Shalev2014}: ``divide and conquer". With a random forest, 
many decision trees are trained on a random subset of the training data 
with a randomly picked subset of dimensions. 
One decision tree alone tends to highly overfit. 
But the random forest, based on the aggregation of the trees and a majority decision rule, 
has very good and consistent results. 
The important idea is to introduce some randomness between the trees.
The concept of a random forest is analogous to extreme pruning successfully utilized by the cryptography community
for sphere decoding~\cite{Gama2010}. They built trees having low success rate and repeated the operation many times with different bases of the lattice. They observed that complexity decreases much faster than the performance deterioration. 
This successful concept was also known in Ordered Statistics Decoding two decades ago.
Therefore, in case of sub-optimality of the network, a solution can be to duplicate the network and introduce randomness instead of increasing the number of parameters in the DNN. An easy way to introduce randomness is to initialize neural networks with distinct $\hat{z}_0$ obtained via different manner.
An instance of such system is illustrated  in Figure~\ref{fig_Blockrepre}. The first DNN is initialized
with a random $\hat{z}_0$, while the second DNN receives an initial $\hat{z}_0$ obtained by ZF.

\section{Training statistics}
Only a limited amount of studies discuss what training statistics should be used for efficient training of a neural-based decoder. In \cite{Gruber2017}, they introduce the notion of Normalized Validation Error (NVE) to investigate which SNR is most suited for efficient training. 
They empirically observed that a SNR neither too high nor too low is the most efficient. In most papers, authors mix noisy data obtained at different SNRs to perform training, in hope that the network is efficient at all those SNRs.
To the best of our knowledge, in all papers on neural networks for decoding, the input message $z$ associated to a noisy received signal $y$ is used as label for the training. \\

Regardless of the noise, the label that should be used for a given $y$ is what would have been decoded by the optimal decoder, not the transmitted sequence.
Consider for instance a simple BPSK. 
If the noise moves a point (e.g. +1) further than the decoding threshold (e.g. -0.2), one should not tell the neural network to try to recover the original point (here +1): it should decode the point associated to the region the received $y$ belongs to (here -1).

Let us call $\mathcal{C}$ a given constellation/code/lattice that we want to train to decode and $c_{i}$ an element of $\mathcal{C}$.
Leaving apart the notion of SNR, the optimal decoder (which we could also call the Voronoi classifier)
performs the following operation: given a $y$ (anywhere) in the space of $\mathcal{C}$, it finds the $c_{i}$  associated to the decoding (Voronoi) region where $y$ is located.
Moreover, if we want the network to learn the entire structure of $\mathcal{C}$, the training sample should be composed of points sampled randomly in its space. Equivalently, one can randomly choose elements of $\mathcal{C}$ (with equiprobability) and add uniformly distributed noise.

Nevertheless, to get quasi-maximum-likelihood decoding (MLD) performance on the Gaussian channel, the network doesn't need to learn the entire structure
of $\mathcal{C}$ but rather the most relevant decision boundaries \textit{around the $c_{i}$}. Indeed, some regions along the boundaries
are so far from $c_i$ such that the Gaussian noise almost never sends $c_i$ to those regions. Therefore, a quasi-MLD network can potentially make many simplifications compared to a perfect MLD network and thus reduce its complexity. These simplifications can be learned by training the network with Gaussian noise. \\

Unfortunately, getting MLD label can be very costly (especially compared to using the input message $z$): any sample should be decoded with the optimal decoder and potentially stored.
Hence, if we were to use $z$ as label for the training due to limited resources, what SNR should be used on the Gaussian channel?  In light of the above discussion, we would want both to learn the necessary structure of the code to get quasi-MLD performance (i.e. the SNR should not be too high) but the ``noise" in the label (i.e. messages that are wrongly labeled w.r.t. the optimal decoder) should not be too high either. Empirically, we observed that the SNR corresponding to an error probability of $10^{-2}$ is a good trade-off (only one sample out of 100 is mis-labeled but the SNR is low enough to properly explore $\mathcal{C}$).


\section{Simulation results}

In this section we present neural networks performance observed under several settings.
For each of these settings the results we report are the best complexity-performance trade-off we obtained, i.e.  we decreased the network size as much as possible while keeping near MLD performance. \\

\begin{figure}
\centering
\includegraphics[angle=-90,width=0.77\columnwidth]{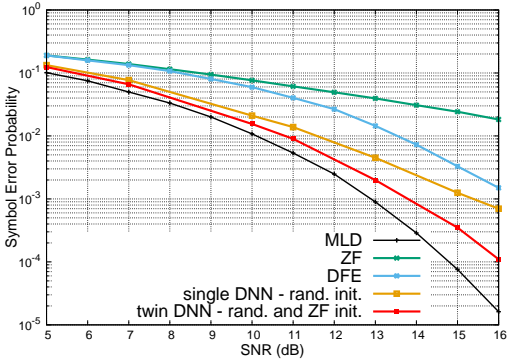}
\caption{First simulations, with \textbf{small batch size} training.}
\label{fig_smallBS}
\vspace{-4mm}
\end{figure}

For the first set of simulations, depicted in Figure \ref{fig_smallBS}, the settings are the following. We take $n=8$ and $M=5$ levels on each $z_{i}$. The MIMO channel is a static channel randomly sampled from an i.i.d. Gaussian matrix.
The considered matrix instance has condition number $17$ and Hermite constant $-4.7$dB (as a real lattice), i.e., this is a bad channel realization and an interesting challenge to our DNN.
Additionally, we used the multilevel activation function. 
The training is done in a regular way with the Adam optimizer and a small batch size ($\approx 200$). The multilevel MIMO detector used for these simulations has $1.25n$ iterations, $\xi$ is of size $7n$ and $v$ of size $n$. Hence, the twin-DNN has $2 \times 1.25 n \times 42n^2 \approx 100n^3$ parameters (which is about 10 times smaller than $5^8$).

We observe that the twin-network DNN performance is close to the MLD performance and clearly outperforms the single DNN (we show only the curve for the randomly initialized single DNN because it matches the one initialized with the ZF point). This means that, under a different initialization, the two single DNNs are almost never wrong at the same time (except for the cases that cannot be recovered by the optimal decoder). Hence, this approach can be beneficial to improve a sub-optimal neural network. \\

\begin{figure}
\centering
\includegraphics[angle=-90,width=0.77\columnwidth]{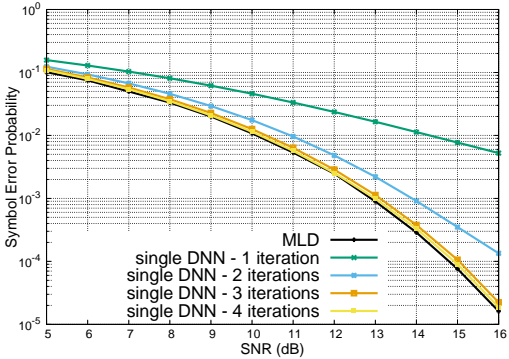}
\caption{Second simulations, with \textbf{large batch size} ($\approx3e^{4}$).}
\label{fig_largeBS}
\vspace{-5mm}
\end{figure}

The second set of simulations was performed under the same settings as the one described above, but the batch size is increased to $\approx 3e^4$ to train the network. Moreover, the size of the $\xi$ layer is decreased to $4n$. In Figure \ref{fig_largeBS}, we show a significant improvement of performance for the single DNN case: within just three iterations ($<<$~1.25n) and with a decreased network size we manage to get near-MLD performance (even though the number of parameters in the network is decreased to $3 \times 24n^2$).
We don't believe that the improvement is caused by a larger amount of data used to train the network: Firstly, in the small-batch simulations we let the networks learn for a large enough amount of time. Secondly, the convergence to quasi-MLD performance with a large batch size is very fast.
We rather believe that a non-noisy gradient is better suited for efficient learning in our settings.\\

\begin{figure}
\centering
\includegraphics[angle=-90,width=0.77\columnwidth]{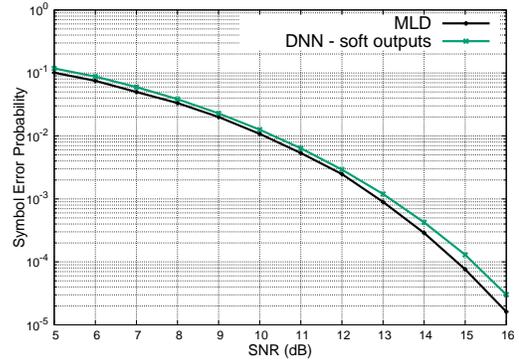}
\caption{DNN with soft outputs.}
\label{fig_softOutputs}
\vspace{-4mm}
\end{figure}

In this work, we also aim at comparing the performance of multilevel activation functions and one-hot encoding. Note that one-hot encoding associated to the soft-max activation function yields soft outputs. Hence, we modify the network used in  Figure \ref{fig_largeBS} by replacing each $M$-level output neuron (i.e. the neurons labeled $z_{k+1}$ in Figure \ref{fig_DNNarchitecture}) by $M$ neurons to get soft outputs. Moreover, we used 10 iterations. The result obtained is depicted in Figure \ref{fig_softOutputs}. We observe that we don't manage to get quasi-optimal performance as in Figure \ref{fig_largeBS}. Additionally, the training phase of this network took significantly more time than the previous one and required much more fine tuning of hyper-parameters. To summarize, this network is more complex and harder to train.\\

\begin{figure}
\centering
\includegraphics[angle=-90,width=0.77\columnwidth]{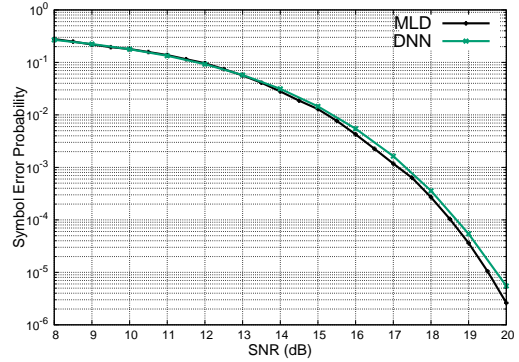}
\caption{DNN for the $T55$ MIMO channel.}
\label{fig_T55}
\vspace{-4.5mm}
\end{figure}

Finally, we perform a last simulation on the $T55$ MIMO channel used in \cite{Samuel2017}. The associated matrix is ill-conditioned, which makes it challenging for linear detectors but not necessarily for the sphere decoder. We take $n=16$, $M=5$ levels with the multilevel activation function on output neurons. We observe in Figure \ref{fig_T55} that this situation is well handled by our neural network. \\

\begin{figure}
\centering
\includegraphics[angle=-90,width=0.77\columnwidth]{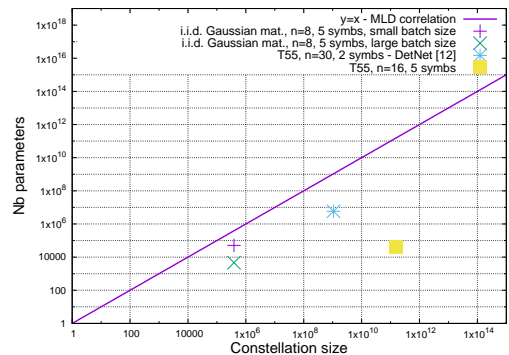}
\caption{Complexity analysis of the considered models.}
\label{fig_complex_ana}
\vspace{-5mm}
\end{figure}

\vspace{-4.5mm} 
The complexity of the different models presented in this section is summarized in Figure \ref{fig_complex_ana}. 
We plot the number of parameters (number of edges) of the network as a function of the cardinality of the constellation (obtained as $M^n$). We also write in blue the complexity of the network used in \cite{Samuel2017} for the T55 MIMO matrix.
We believe that the number of parameters indicated in blue could be diminished without degrading the performance if a larger batch size is used in training. \\

\begin{figure}
    \centering
    \includegraphics[angle=-90,width=0.77\columnwidth]{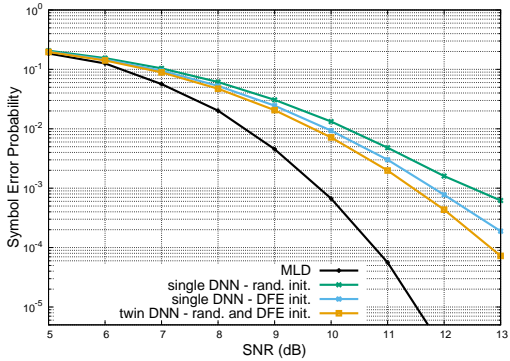}
     \caption{MIMO channel is taken to be the generator of $BW_{16}$.}
     \label{fig_BW16}
\vspace{-4mm}
\end{figure}

In light of these results, we may conclude that deep learning, with the proposed approach, is competitive for a large range of MIMO channels. However, deep learning in some extremal situations is difficult to set up, namely for specific channels where the function to be approximated is very challenging.
For instance, if the MIMO channel is the generator matrix of a dense lattice (e.g. $E_{8}$, $BW_{16}$, $\Lambda_{24}$ \cite{Conway1999}), the function to learn is  more complex (see next section) and even a neural network with a large number of iterations and an increased size for each layer fails to achieve near-MLD performance, as shown in Figure \ref{fig_BW16}. Fortunately these extremal communication channels are rarely encountered.

\section{Connection with (infinite) lattice decoding}

Lattice modeling of the MIMO channel is not always successful because of the finite number of levels which induces a finite constellation: the MLD point in the lattice can be out of the finite MIMO constellation. With the regular sphere decoder, it is possible to bound the number of states that each component of $z$ can take and overcome this issue. However, if complexity reduction techniques are used as preprocessing, such as basis reduction, then this issue is difficult to avoid. Similarly, the hyperplane logical decoder (HLD) introduced in \cite{Corlay2018}, a neural network based lattice decoder, cannot be used (i.e. leads to disappointing performance) for MIMO detection because it can detect messages which are not in the finite constellation. \\

\vspace{-2mm}
In this section, we present a new strategy to avoid this issue while using a lattice-based approach. Namely, we show how the detection can be performed in the fundamental parallelotope $\mathcal{P}$,
given a quasi-Voronoi-reduced lattice basis (see \cite{Corlay2018}),
 and still detect only possible messages belonging to the finite alphabet. This leads to both:
\begin{itemize}
\item A better understanding of the hardness of the problem that the neural network should solve.
\item A new strategy for lattice-based multilevel MIMO detection with neural networks. 
\end{itemize}

\begin{figure}
\centering
\vspace{-2mm}
\includegraphics[width=0.6\columnwidth]{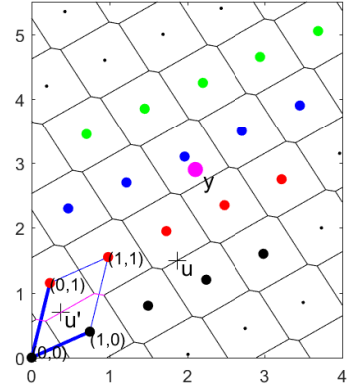}
\vspace{-4mm}
\caption{Example of lattice-based MIMO detection in $\mathcal{P}$.}
\label{fig_para}
\vspace{-5mm}
\end{figure}

We present the approach in four steps. Consider that the $n$-th component of $z$ is to be detected.
\begin{itemize}
\item Step 1: Go in the fundamental parallelotope $\mathcal{P}$ and consider only the $n-1$ first coordinates of $y$.
\begin{align*}
\vspace{-1mm}
y'^{[n-1]} & = y^{[n-1]} \ \text{mod} \ \mathcal{P}^{[n-1]} = y^{[n-1]} -  (tG)^{[n-1]},
\vspace{-1mm}
\end{align*}
where  $t = \lfloor x G^{-1} \rfloor$. 

\item Step 2: Compute the decision boundary function (in pink on Figure~\ref{fig_para}): 
$
u' = g(y'^{[n-1]}).
$
\item Step 3: Go back to the original location.
\begin{align*}
u = u'  + \Sigma_{i}t_{i} \frac{b_{i} \cdot e_{n}}{||e_{n}||^{2}},
\end{align*}
where $\{b_i\}$ is the lattice basis and $\{e_i\}$ defines the coordinate system.
\item Step 4: Apply the multilevel sigmoid function on $u-y_{n}$ with delays equal to:
\vspace{-1mm}
\begin{align*}
\tau = \frac{e_{n}\cdot b_{n}}{||e_{n}||^{2}}.
\end{align*}
\end{itemize}
The main operational cost of this algorithm is due to the decision boundary function.
It is closely related to the Boolean equation of the HLD and can be computed with a DNN.

\end{document}